\input harvmac

\input amssym.tex


\lref\ChandiaIX{
  O.~Chandia,
  ``A note on the classical BRST symmetry of the pure spinor string in a curved
  background,''
  JHEP {\bf 0607}, 019 (2006)
  [arXiv:hep-th/0604115].
}

\lref\GuttenbergIC{
  S.~Guttenberg,
  ``Superstrings in General Backgrounds,''
  arXiv:0807.4968 [hep-th].
}

\lref\BerkovitsRJ{
  N.~Berkovits and C.~Vafa,
  ``Towards a Worldsheet Derivation of the Maldacena Conjecture,''
  JHEP {\bf 0803}, 031 (2008)
  [AIP Conf.\ Proc.\  {\bf 1031}, 21 (2008)]
  [arXiv:0711.1799 [hep-th]].
}

\lref\BerkovitsBT{
  N.~Berkovits,
  ``Pure spinor formalism as an N = 2 topological string,''
  JHEP {\bf 0510}, 089 (2005)
  [arXiv:hep-th/0509120].
}

\lref\ToninMM{
  M.~Tonin,
  ``Pure Spinor Approach to Type IIA Superstring Sigma Models and Free
  Differential Algebras,''
  arXiv:1002.3500 [hep-th].
}

\lref\BerkovitsNekrasov{
  N.~Berkovits and N.~Nekrasov,
  ``Multiloop superstring amplitudes from non-minimal pure spinor formalism,''
  JHEP {\bf 0612}, 029 (2006)
  [arXiv:hep-th/0609012].
}

\lref\BedoyaAV{
  O.~A.~Bedoya, D.~Z.~Marchioro, D.~L.~Nedel and B.~C.~Vallilo,
  ``Quantum Current Algebra for the $AdS_5 \times S^5$ Superstring,''
  arXiv:1003.0701 [hep-th].
}

\lref\BedoyaIC{
  O.~A.~Bedoya and O.~Chandia,
  ``One-loop conformal invariance of the type II pure spinor superstring in a
  curved background,''
  JHEP {\bf 0701}, 042 (2007)
  [arXiv:hep-th/0609161].
}

\lref\PulettiVB{
  V.~G.~M.~Puletti,
  ``Operator product expansion for pure spinor superstring on AdS(5) x S**5,''
  JHEP {\bf 0610}, 057 (2006)
  [arXiv:hep-th/0607076].
}

\lref\BianchiIM{
  M.~Bianchi and J.~Kluson,
  ``Current algebra of the pure spinor superstring in AdS(5) x S(5),''
  JHEP {\bf 0608}, 030 (2006)
  [arXiv:hep-th/0606188].
}

\lref\metsaev{R. Metsaev and A. Tseytlin, 
{\it Type
IIB Superstring Action in $AdS_5\times S^5$ Background,}
Nucl. Phys. B533 (1998) 109, hep-th/9805028.}

\lref\OdaZM{
  I.~Oda and M.~Tonin,
``On the Berkovits covariant quantization of GS superstring,''
  Phys.\ Lett.\  B {\bf 520}, 398 (2001)
  [arXiv:hep-th/0109051].
}

\lref\MatoneFT{
  M.~Matone, L.~Mazzucato, I.~Oda, D.~Sorokin and M.~Tonin,
  ``The superembedding origin of the Berkovits pure spinor covariant
  quantization of superstrings,''
  Nucl.\ Phys.\  B {\bf 639}, 182 (2002)
  [arXiv:hep-th/0206104].
}

\lref\ValliloMH{
  B.~C.~Vallilo,
  ``One loop conformal invariance of the superstring in an AdS(5) x S(5)
  background,''
  JHEP {\bf 0212}, 042 (2002)
  [arXiv:hep-th/0210064].
}

\lref\BerkovitsXU{
  N.~Berkovits,
  ``Quantum consistency of the superstring in AdS(5) x S**5 background,''
  JHEP {\bf 0503}, 041 (2005)
  [arXiv:hep-th/0411170].
}

\lref\MazzucatoFV{
  L.~Mazzucato and B.~C.~Vallilo,
  ``On the Non-renormalization of the AdS Radius,''
  JHEP {\bf 0909}, 056 (2009)
  [arXiv:0906.4572 [hep-th]].
}

\lref\BerkovitsUE{
  N.~Berkovits and P.~S.~Howe,
  ``Ten-dimensional supergravity constraints from the pure spinor formalism
  for the superstring,''
  Nucl.\ Phys.\  B {\bf 635}, 75 (2002)
  [arXiv:hep-th/0112160].
}

\lref\GuttenbergIC{
  S.~Guttenberg,
  ``Superstrings in General Backgrounds,''
  arXiv:0807.4968 [hep-th].
}

\lref\BerkovitsRJ{
  N.~Berkovits and C.~Vafa,
  ``Towards a Worldsheet Derivation of the Maldacena Conjecture,''
  JHEP {\bf 0803}, 031 (2008)
  [AIP Conf.\ Proc.\  {\bf 1031}, 21 (2008)]
  [arXiv:0711.1799 [hep-th]].
}

\lref\BerkovitsYR{
  N.~Berkovits and O.~Chandia,
  ``Superstring vertex operators in an AdS(5) x S(5) background,''
  Nucl.\ Phys.\  B {\bf 596}, 185 (2001)
  [arXiv:hep-th/0009168].
}


\def\a{{\alpha}}
\def\b{{\beta}}

\def\l{{\lambda}}
\def\t{{\theta}}

\def\p{\partial}

\def\half{{1\over 2}}

\def\Tb{{\bf{T}}}

\def\ah{{\hat\alpha}}
\def\bh{{\hat\beta}}
\def\lh{{\hat\lambda}}
\def\wh{{\hat w}}
\def\gh{{\hat\gamma}}
\def\dh{{\hat\delta}}

\def\tr{{\rm Tr}\,}

\def\a{\alpha}
\def\b{\beta}
\def\g{\gamma}
\def\d{\delta}

\def\l{\lambda}

\def\p{\partial}
\def\t{{\theta}}

\def\ah{{\hat \alpha}}


\Title{\vbox{\vskip 0.5cm
}}
 {\vbox{{}\medskip\medskip\centerline{Taming the $b$ antighost
with Ramond-Ramond
flux}}}
\smallskip
\centerline{Nathan Berkovits${}^a$ and Luca Mazzucato${}^b$}
\medskip
\centerline{\it ${}^a$ Instituto de F\'{\i}sica Te\'orica, UNESP -
Universidade Estadual Paulista}
\smallskip
 \centerline{\it S\~ao Paulo, SP 01140-070, Brasil}
\medskip
\centerline{\it ${}^b$ Simons Center for Geometry and Physics, 
SUNY at Stony Brook}
\smallskip
 \centerline{\it Stony Brook, NY 11794-3840, USA}
\medskip

\bigskip
\vskip 0.5cm

\noindent In the pure spinor formalism for the superstring, the 
$b$ antighost is necessary for multiloop amplitude computations and is 
a composite operator constructed to satisfy $\{Q,b\}=T$ where $Q$ is
the BRST operator and $T$ is the holomorphic stress-tensor. In superstring 
backgrounds with
only NS-NS fields turned on, or in flat space, 
one needs to introduce ``non-minimal'' variables in
order to construct the $b$ antighost. However,
in Type II backgrounds where the Ramond-Ramond bispinor field-strength satisfies certain conditions, the $b$ antighost 
can be constructed
without the non-minimal variables. Although the $b$ antighost in these
backgrounds is not holomorphic, its antiholomorphic derivative is BRST-trivial.
We discuss the properties of this operator both in the $AdS_5\times S^5$ 
background and in a generic curved background.
\centerline{}
{\noindent
}
\vskip 0.5cm

\noindent

 \vskip 0.5cm

\Date{April 2010}


\newsec{Introduction}

Over the last ten years, the pure spinor formalism for the superstring
has been used successfully to compute superstring scattering amplitudes
(see \ref\bedoya{N. Berkovits and O. Bedoya, ``GGI lectures on the
pure spinor formalism of the superstring,''
[arXiv:0910.2254].} for a recent review).
A major advantage over computations using
the Ramond-Neveu-Schwarz formalism is that
spacetime supersymmetry is manifest in the pure spinor formalism
so one does not need to sum over
spin structures to see cancellations in the multiloop amplitudes.

Since $(b,c)$ and $(\hat b,\hat c)$
reparameterization ghosts are not fundamental worldsheet
variables in the
pure spinor formalism, $g$--loop scattering amplitudes ${\cal A}_g$ are defined as
in topological string theory where the left-moving
$b$ antighost and right-moving $\hat b$ antighost are composite fields
constructed to satisfy 
$$
\{Q,b\}=T, \quad \{\hat Q,\hat b\}=\hat T \ ,
$$ 
where $Q$ and $\hat Q$ are the left and right-moving BRST operators and
$T$ and $\hat T$ are the left and right-moving stress tensors. As in topological
string theory, the integration measure
is then defined by contracting $(3g-3)$ composite $b$ antighosts with the
Beltrami differentials $\mu$ corresponding to the $(3g-3)$ 
Teichmuller moduli $\tau$ of
the genus $g$ Riemann surface
\eqn\ampli{
{\cal A}_g=\int d^{3g-3}\tau \int d^{3g-3}\bar\tau\langle 
(\int \mu b)^{3g-3}(\int \bar\mu \hat b)^{3g-3}\prod_{i=1}^N 
\int d^2 z_i V_i(z_i) \rangle \ .
}

In a flat background, the construction of the $b$ antighost 
satisfying $\{Q,b\}=T$
is complicated and requires the introduction of non-minimal worldsheet
variables. In the ``minimal'' pure spinor formalism, one has the
usual $(x^m,\t^\a,\hat\t^\ah)$ Type II superspace variables as well as
the 
left and right-moving bosonic pure
spinor ghosts $(\l^\a,\lh^\ah)$ satisfying $\l\g^m\l=\lh\g^m\lh=0$,
where $\a=1$ to 16 and $\ah=1$ to 16 are ten-dimensional spinor indices 
which
have the same chirality for Type IIB and opposite chirality for Type IIA.
To construct the $b$ antighost in a flat background, one then needs to include
non-minimal variables consisting of a new set of left and
right-moving bosonic pure spinors, 
$(\overline\l_\a,\hat{\overline\l}_\ah)$, as well as a set of left and
right-moving constrained fermions, $(r_\a,\hat r_\ah)$. 
These non-minimal variables satisfy the constraints
$\overline\l\g^m\overline\l=\hat{\overline\l}
\g^m\hat{\overline\l}=0$ and $\overline\l\g^m r 
= \hat{\overline\l}\g^m\hat r=0$,
so that there are an equal number of non-minimal
bosonic and fermionic degrees of freedom.
After modifying the
BRST operator to include the standard non-minimal term,
these new variables decouple from the cohomology
and the physical spectrum.

In addition to allowing construction of a $b$ antighost satisfying $\{Q,b\}=T$,
these non-minimal variables also allow functional integration over the pure
spinor ghosts, where $\overline\l_\a$ is interpreted as the complex conjugate
of $\l^\a$ and $\hat{\overline\l}_\ah$ is interpreted as the complex conjugate
of $\lh^\ah$. Although scattering amplitudes have been computed
using this prescription
only in a flat background, it is natural to ask if this 
construction of the $b$ antighost
generalizes to curved supergravity backgrounds. One interesting background 
to consider is the $AdS_5\times S^5$ background with Ramond-Ramond flux.

In a recent paper \ref\extend{N. Berkovits, ``Simplifying and extending
the $AdS_5\times S^5$ pure spinor formalism,''
JHEP 0909:051,2009, arXiv:0812.5074.} by one of the authors, 
it was argued that unlike in a flat background, non-minimal
variables are not needed to construct the $b$ antighost in an $AdS_5\times S^5$
background. Instead of introducing new non-minimal variables $({\overline\l}_\a,
\hat{\overline\l}_\ah)$ to play the role of the complex conjugates of 
$(\l^\a,\lh^\ah)$, one can simply define 
\eqn\relnm{\overline\l_\a \equiv\g^{01234}_{\a\ah}
\lh^\ah, \quad \hat{\overline\l}_\ah \equiv \g^{01234}_{\a\ah} \l^\a,}
where $\g^{01234}_{\a\ah}$ is the five-form gamma matrix in the direction
of the five-form Ramond-Ramond flux.
So after multiplying by $\g^{01234}$, the original 
left and right-moving pure spinor
ghosts can be interpreted as complex conjugates of each other. In a flat
background, this interpretation
is not possible since
$\l^\a \overline\l_\a = \g^{01234}_{\a\ah}\l^\a\lh^\ah$ is BRST-trivial, 
so it cannot be interpreted
as a positive-definite quantity. But in an $AdS_5\times S^5$ background,
$\g^{01234}_{\a\ah}\l^\a\lh^\ah$ is in the BRST cohomology: it is the
vertex operator for
the radius modulus. So it is consistent to interpret
$\g^{01234}_{\a\ah}\l^\a\lh^\ah$ as a positive-definite quantity
since it cannot be gauged away. After interpreting the complex conjugate
of the pure spinor variables as in \relnm,
the construction of the $b$ antighost in an $AdS_5\times S^5$ background is
straightforward.

In the first part of this paper, this 
construction of the $b$ antighost in an $AdS_5\times S^5$
background will be shown to satisfy the necessary properties for
consistency of the amplitude prescription of \ampli. In 
addition to satisfying $\{Q,b\}=T$, it will be shown that
the $b$ antighost also satisfies $\{\hat Q, b\}=0$.
However, unlike the left-moving $b$ antighost in a flat background, the 
$b$ antighost in an $AdS_5\times S^5$ background is not holomorphic, i.e.
it does not satisfy $\bar\partial b=0$. Instead it satisfies 
\eqn\nothol{\bar\partial b = [\hat Q,{\cal O}]}
where ${\cal O}$ is defined
by taking the antiholomorphic contour integral 
of $\hat b$ around $b$.  One similarly finds that
the $\hat b$ antighost 
is not antiholomorphic  
and instead satisfies
\eqn\notantihol{\partial \hat b = [ Q,\hat{\cal O}]}
where $\hat {\cal O}$ is defined by taking the holomorphic contour
integral of $b$ around $\hat b$.

To prove \nothol, one uses the properties
\eqn\proper{\{Q,b\}=T,\quad \{\hat Q, b\}=0,\quad \{\hat Q,\hat b\}=\hat T,
\quad\{Q,\hat b\}=0}
to show that 
\eqn\showt{\bar\partial b = [\hat T_{-1}, b] = [\{\hat Q, \hat b_{-1}\}, b] 
= [\hat Q, {\cal O}]}
where $[\hat T_{-1}, X]$ and 
$\{\hat b_{-1}, X\}$ denote the antiholomorphic
contour integral of $\hat T$ and $\hat b$ 
around $X$, and ${\cal O} \equiv \{\hat b_{-1}, b\}$. 
One can similarly use \proper\ to prove \notantihol\ where
$\hat{\cal O} \equiv \{ b_{-1},\hat b\}$ and $\{b_{-1},X\}$ denotes
the holomorphic contour integral of $b$ around $X$.

Although this non-holomorphic structure of the $b$ and $\hat b$
antighosts is unusual, \nothol\
and \notantihol\ 
should be enough
for consistency of the amplitude prescription of \ampli.
In order that $\int \mu b$ in \ampli\
is invariant under the shift $\mu\to\mu+\bar\partial\nu$ for any $\nu$, 
one usually requires that $\bar\partial b=0$. However, if one can ignore
surface terms coming from the boundary of Teichmuller moduli space,
it is sufficient to require the milder condition 
\eqn\milder{ \bar\partial b=[\hat Q,{\cal O}] \ .  } This can be shown
by pulling $\hat Q$ off of 
${\cal O}$ and using $[\hat Q,V]=\{\hat Q,b\}=0$ and $\{\hat Q,\hat b\}=\hat
T$ to obtain terms
which are total derivatives in the Teichmuller moduli. If one can ignore
surface terms from the boundary of moduli space, these 
total derivatives do not contribute. For backgrounds such
as $AdS_5\times S^5$ which preserve
spacetime supersymmetry, one does not
expect the integrand of the scattering amplitude
to diverge near the boundary of moduli space, so it should be OK
to ignore these surface terms. However, the role of such terms in the $AdS_5\times S^5$ Ramond-Ramond
background deserves 
further investigation.

In the second part of the paper, we show that a similar construction
of the $b$ antighost is possible whenever the supergravity background includes
a Ramond-Ramond field strength which, when expressed in bispinor
notation as $P^{\a\bh}$, obeys certain conditions. In the type II superstring, the dependence of the superfield $P$ on the Ramond-Ramond $p$-form field strengths $F_p$ is
\eqn\rrsuperfield{\eqalign{
{\rm IIB:}\qquad{1\over {g_s}} P=&\gamma^{a_1}F_{a_1}+{1\over3!}\gamma^{a_1a_2a_3}F_{a_1a_2a_3}+{1\over2\cdot 5!}\gamma^{a_1\ldots a_5}F_{a_1\ldots a_5} \ , \cr
{\rm IIA:}\qquad{1\over {g_s}} P=& F_0+{1\over2!}\gamma^{a_1a_2}F_{a_1a_2}+{1\over 4!}\gamma^{a_1\ldots a_4}F_{a_1\ldots a_4} \ .
}}
For example, in the type IIB $AdS_5\times S^5$ background, $P^{\a\ah} = 
\g_{01234}^{\a\ah}$ whose inverse is $(P^{-1})_{\a\ah}=\g^{01234}_{\a\ah}$.
We will show that when the R-R superfield is covariantly constant
and invertible, 
the state $(P^{-1})_{\a\ah}\l^\a\lh^\ah$ is in
the BRST cohomology and one can redefine 
\eqn\relnmg{\overline\l_\a \equiv (P^{-1})_{\a\ah}
\lh^\ah, \quad \hat{\overline\l}_\ah \equiv (P^{-1})_{\a\ah} \l^\a,}
and construct a $b$ antighost such that $\{Q,b\}=T$. In backgrounds where the Ramond-Ramond field strength
satisfies the additional requirement
\eqn\conditionP{
P\gamma^kP=f_m^k\gamma^m \ ,
}
where $f_m^k$ is a superfield, 
it will be shown that \proper\ is still satisfied, 
which implies \nothol\ and \notantihol. We will solve the condition \conditionP\ explicitly in terms of the type II R-R fluxes. 
Surprisingly, the construction of the $b$ antighost in a curved
NS-NS background is more complicated than in a R-R background since
it requires non-minimal variables.

The fact that non-minimal variables in the pure spinor formalism
are not necessary
in backgrounds where the Ramond-Ramond superfield is covariantly constant 
and invertible
should have consequences for scattering amplitudes
in these backgrounds. Firstly, it would be interesting to know if there
are any examples of such backgrounds besides $AdS_5\times S^5$.
Since non-minimal variables play an important
role in the proof of non-renormalization theorems
in a flat background, it would be interesting to study 
non-renormalization theorems in these Ramond-Ramond backgrounds.

Another interesting feature of this paper is the
construction of ${\cal O}$ in \nothol\ in 
terms of the single pole between the $b$ and $\hat b$ antighost. 
We are not aware of any previous discussion of such a construction,
and there should be a natural geometrical interpretation of ${\cal O}$
in backgrounds where $b$ is not holomorphic but \proper\ is satisfied.


\newsec{$AdS_5\times S^5$ Background}

Superstring propagation in the $AdS_5\times S^5$ background is described by a non-linear sigma model defined on the supercoset $PSU(2,2|4)/SO(1,4)\times SO(5)$. To set the notation we briefly collect some facts about the pure spinor sigma model. 

A coset representative $g(\sigma)$ transforms as $g'(\sigma)=g_0 g(\sigma)h(\sigma)$, where $g_0$ is an element of the global $PSU(2,2|4)$ and $h(\sigma)$ is an element of the local $SO(1,4)\times SO(5)$ Lorentz group. The left-invariant currents $J=g^{-1}dg$ can be decomposed according to the $Z_4$ automorphism of the super Lie algebra $PSU(2,2|4)$ as
\eqn\notat{J_0 = (g^{-1}\p g)^{[ab]} \Tb_{[ab]},\quad
J_1 = (g^{-1}\p g)^{\a} \Tb_{\a},\quad
J_2 = (g^{-1}\p g)^{a} \Tb_{a},\quad
J_3 = (g^{-1}\p g)^{\ah} \Tb_{\ah},}
where $\Tb_A$ are the super Lie algebra generators. They satisfy the Maurer-Cartan equations 
\eqn\mceq{
\partial \bar J+\bar\partial J+ [ J,\bar J]= 0 \ ,
}
which can be conveniently split according to the $Z_4$ grading. We will need the left and right-moving ghosts and their conjugate momenta $(\l^\a,w_\a)$ and $(\hat\l^\ah,\hat
w_\ah)$. As anticipated in \relnmg, it will be convenient to redefine the hatted worldsheet quantities by introducing a factor of the constant Ramond-Ramond superfield $P^{\a\ah}=(\g_{01234})^{\a\ah}$ and its inverse $P_{\a\ah}
=(\g_{01234})_{\a\ah}$ 
\eqn\redefinition{
\hat\lambda_\a\equiv P_{\a\hat\a}\hat\l^{\hat\a} \ , \quad \hat w^\a\equiv P^{\a\ah}\hat w_\ah  \ ,\quad
(J_3)_\a\equiv P_{\a\hat\a}J_3^{\hat\a}  .}
The worldsheet action reads
\eqn\actionads{\eqalign{
S=&{R^2\over2\pi}\int d^2z\Bigl(\half\eta_{ab}J^a\bar J^b+{3\over4}(J_3)_\a\bar J_1^\a-{1\over 4}J_1^\a\bar (J_3)_\a \cr &+w_\a(\bar\nabla\l)^\a+\hat w^\a(\nabla\hat\l)_\a-\half\eta_{[ab][cd]} N^{ab}N^{cd}\Bigr) \ ,
}}
where $N^{ab}$ and $\hat N^{ab}$ are the $SO(1,4)\times SO(5)$ Lorentz generators of the pure spinors and $\eta_{[ab][cd]}=(\eta_{\tilde a[\tilde c}\eta_{\tilde d]\tilde b},-\delta_{\bar a[\bar c} \delta_{\bar d]\bar b})$, where $\tilde a=0,\ldots,4$ and $\bar a=5,\ldots,9$ are the $AdS_5$ and $S^5$ directions respectively. We introduced the covariant derivatives
$$(\nabla\l)^\a  = \partial  \l^\a+ \half J_0^{ab}(\gamma_{ab}\l)^\a \ ,\quad (\bar\nabla\hat\l)_\a  = \bar\partial  \hat\l_\a- \half J_0^{ab}(\gamma_{ab}\hat\l)_\a  \ .$$

The physical states are vertex operators in the cohomology of the nilpotent BRST charge $Q+\hat Q$
\eqn\brstch{
Q=\oint d\sigma \l^\a (J_3)_\a \ ,\qquad 
\hat Q=\oint d\sigma \hat\l_\a\bar J_1^\a \ ,
}
that generate the following BRST transformations \BerkovitsYR\
\eqn\brsttr{\eqalign{
Q J_1^\a=(\nabla\l)^\a \ ,\quad Q J_2^a=(\l\gamma^aJ_1) \ ,\quad Q (J_3)_\a=-(\l\gamma_a)_\a J_2^a \ ,
}}
$$
\hat Q J_1^\a=-(\gamma_a\hat\l)^\a J_2^a \ ,\quad \hat Q J_2^a=(\hat \l\gamma^aJ_3) \ ,\quad \hat Q (J_3)_\a=(\bar\nabla\hat\l)_\a \ ,
$$
$$
Q w_\a=(J_3)_\a \ , \quad \hat Q \hat w^\a=-\bar J_1^\a \ ,\quad \hat Q w_\a=Q\hat w^\a=0 \ ,
$$
$$
Q N^{ab}=\half (J_3\gamma^{ab}\l) \ ,\quad \hat Q\hat N^{ab}=\half(\hat\l\gamma^{ab}\bar J_1) \ .
$$
In terms of the $PSU(2,2|4)$ super Lie algebra, the grading one and the grading three subspaces are related by hermitian conjugation which implies
$$
(\l^\a)^\dagger=\hat\l_{\a} 
 \ .
 $$

The stress tensor of the worldsheet theory is 
\eqn\stress{\eqalign{
T
=&-\half J_2^a J_2^b\eta_{ab}+J_1^\a J_{3\a}-w_\a\nabla\l^\a \ ,
}}
and it is easy to check that it satisfies $\{Q,T\}=\{\hat Q,T\}=0$. The consistency of the
theory at the quantum level has been checked in \ValliloMH\BerkovitsXU\MazzucatoFV.

\subsec{The antighost}

Before we consider the $b$ antighost, let us take a quick detour and introduce a useful projection operator. The conjugate momentum to the pure spinor variable, that we denoted $w$, may only appear in expressions that are gauge invariant with respect to the local symmetry
\eqn\gauge{
\delta_w w_\alpha=(\gamma^a\lambda)_\alpha  \Lambda_a\ ,
}
which is generated by the pure spinor constraint. As
in flat space, the only gauge invariant combinations of $w$ 
are the $SO(1,9)$ Lorentz generators $N^{ab}$ and 
the ghost number current $J_{gh}$. However, instead of
working with $N^{ab}$ and $J_{gh}$, it will be convenient
to define a projection operator $(1-K)^\a_\b$ which selects
out the gauge-invariant components of $w_\a$. In other words,
$(1-K)^\a_\b \d_w w_\a=0$ under \gauge. 

Consider the following projection operator,\foot{A brief historical 
comment. The projection operator $K_\a^\b$ was first introduced in 
\OdaZM\MatoneFT\ in a flat background, in the context of a 
semiclassical derivation of the pure spinor formalism from a 
Green-Schwarz type action. However, the $\lh$ variable in
\OdaZM\MatoneFT\ is a fixed spinor so
the formalism in \OdaZM\MatoneFT\ is not manifestly Lorentz 
covariant, but is only valid in a patch of the pure spinor 
manifold. The Lorentz variation of the non-covariant $b$ antighost 
constructed in \MatoneFT\ is BRST exact.}  built out of the inverse 
power of $(\l\hat\l)\equiv P_{\a\ah}\l^\a\lh^\ah$
\eqn\projector{
K^\a_\b={1\over 2(\l\hat\l)}(\gamma^a\hat\l)^\a(\l\gamma_a)_\beta ={1\over 2(\l\hat\l)}(\gamma^a\l)_\beta(\hat\l\gamma_a)^\a\ ,
}
with the following properties
\eqn\kappapro{\eqalign{
(1-K)\gamma^a\lambda=0 \ ,\qquad K\gamma^a\gamma^b\l=0 \ ,\qquad K\nabla\l=0 \ ,\cr
(1-K)\gamma^a\hat\lambda=0\ ,\qquad 
K\gamma^a\gamma^b\hat\l=0 \ ,\qquad K\nabla\lh= 0 \ ,}}
and its traces over the spinor indices are $\tr K=5$ and $\tr(1-K)=11$. By means of the projector $K_\a^\b$ we can introduce the new quantity
\eqn\newgauge{
w_\a(1-K)^\a_\b \ ,
}
which is invariant under \gauge. 

In a flat background, one can construct a similar $K_\a^\b$ by replacing
$\lh_\a$ with the non-minimal variable $\bar\l_\a$. If one interprets 
$\bar\l_\a$
as the complex conjugate of $\l^\a$, $(\l\bar\l)^{-1}$ is formally well-defined
after the point $\l^\a=0$ for all sixteen components is removed
from the theory. However, as discussed in \BerkovitsBT\BerkovitsNekrasov,
there are problems if the negative powers of $(\l\bar\l)$ accumulate
beyond 11. In an $AdS_5\times S^5$ background,
one expects similar problems if the negative powers
of $(\l\lh)$ accumulate beyond a certain amount. However,
as in a flat background,
we expect that our construction of the 
$b$ antighost does not contain enough negative powers of $(\l\lh)$ to cause
problems.

After using
the ten dimensional identity
\eqn\tenid{
(\gamma_{ab})_\alpha{}^\beta(\gamma^{ab})_\gamma{}^\delta=4(\gamma_a)^{\beta\delta}(\gamma_a)_{\alpha\gamma}-2 \delta_{\alpha}^\beta\delta_\gamma^\delta-8\delta_\alpha^\delta\delta_\gamma^\beta \ ,}
the expression for the $AdS_5\times S^5$
antighost in \extend\ can be written in terms of the
$(1-K)^\a_\b$ projector as 
\eqn\bghost{\eqalign{
b=&{(\hat\lambda\gamma_a J_3)J_2^a\over 2(\lambda\hat\lambda)}-w_\a(1-K)^\a_\b J_1^\b\ .
}}
Using the BRST transformations in \brsttr, it was shown in \extend\
that $\{Q,b\}=T$. Note that as in \extend, we will be ignoring possible
normal-ordering corrections to the $b$ antighost throughout this paper
and will only be considering the terms in $b$ which are lowest order in $\a'$.

The other crucial property of the $b$ antighost is
$$
\{\hat Q, b\}= 0 .
$$
Let us prove it. The variation of the first term in \bghost\ is
$$
{1\over 2(\l\hat\l)}\left((\hat\l\gamma_a \bar\nabla\hat\l)J_2^a-(\hat\l\gamma_a J_3)(\hat\l\gamma^a J_3)\right) \ ,
$$
which vanishes because of the pure spinor constraint and the properties of ten dimensional gamma matrices. The variation of the second term in \bghost\ is
$$
w_\a(1-K)^\a_\b(\gamma_a\hat\l)^\b J_2^a \ ,
$$
which vanishes due to the properties of the projector \kappapro.

An analogous construction carries over to the right-moving sector.
The right-moving stress tensor and antighost are
\eqn\tleft{\eqalign{
\hat T
=&-\half\bar J_2^a\bar J_2^b\eta_{ab}+\bar J_1^\a\bar J_{3\a}-\wh\bar\nabla\lh \ ,
}}
\eqn\bleft{\eqalign{
\hat b
=&-{1\over2\l\lh}(\l\gamma_a\bar J_1)\bar J_2^a-\wh^\a(1-K)_\a^\b (\bar J_3)_\b \ .
}}
One can check that $\{\hat Q,\hat b\}=\hat T$ and
$\{Q,\hat b\}=0$.

\subsec{Conservation of the antighost}

We can apply the argument given in the introduction to show that the $b$ antighost is conserved up to BRST exact terms. Let us rewrite \bghost\ in the convenient form
\eqn\bG{\eqalign{b=&{\hat\lambda_\a\over (\lambda\hat\lambda)}G^\alpha \cr
G^{\alpha}=&-\half(\gamma_aJ_3)^\a J_2^a-\lambda^\a (wJ_1)+\half (\gamma^a w)^\a(\lambda\gamma_a J_1) \ .
}}
Since the $b$ antighost is a Lorentz scalar, we have that $\bar\partial b=\bar\nabla b$ and 
\eqn\debi{
\bar\nabla b=\bar \nabla \left({\hat\lambda_\a\over (\lambda\hat\l)}\right) \left(-\half(\gamma_aJ_3)^\a J_2^a+\half (\gamma^a w)^\a(\lambda\gamma_a J_1)\right)+{\hat\lambda_\a\over (\lambda\hat\l)}\bar \nabla G^\a \ .
}
Let us look at the second term in \debi. By using the equations of motion of the action \actionads\ and the Maurer-Cartan equations \mceq\ we find
\eqn\eomsg{\eqalign{
{\hat\lambda_\a\over (\lambda\hat\l)}\bar \nabla G^\a=&b_0+b_w+b_{w\hat w}+b_{ww} \ ,\cr
b_0=&{1\over2(\l\hat\l)}(\hat\l\gamma_a J_3)(J_3\gamma^a\bar J_3)\ ,\cr
b_w=&[w(1-K)\gamma_a]^\a\left( (J_3)_\a\bar J_2^a-(\bar J_3)_\a J_2^a\right) \cr
&+{1\over 2(\l\hat\l)}\left(\half (\bar J_3\gamma_{ab}\gamma_c\hat\l)N^{ab}J_2^c+2(\hat\l\gamma_a J_3) (\bar J_2)_bN^{ab}\right) \ ,\cr
b_{ww}=& -\half[w(1-K)\gamma_{ab}\bar J_1]]N^{ab} \ ,
}}
where the subscript indicates the number of $w$'s and $\wh$'s present in each term. The term $b_{w\hat w}$ is proportional to $\eta_{[ab][cd]}\hat N^{ab} (\hat\l\gamma^{cd})_\a$ which vanishes on the pure spinor constraint.

Let us show that $\bar \nabla b$ is BRST exact. Consider the operator ${\cal O}$ of weight $(2,1)$, defined as the coefficient of the single pole in the OPE of the hatted and unhatted antighosts  
\eqn\method{
\hat b(z,\bar z) b(0)=\ldots+{{\cal O}_{zz\bar z }(0)\over \bar z}+\ldots \ .
}
Since $\{\hat Q,\hat b\}=\hat T$ and $\{\hat Q , b\}= 0$,
by applying $\hat Q$ to \method\ we conclude that
\eqn\hatqb{
\{\hat Q,{\cal O}\}=\bar\nabla b \ .
}

Since the pure spinor superstring in $AdS_5\times S^5$ is an interacting two--dimensional conformal field theory, the OPE \method\ has to be computed in the worldsheet perturbation theory. In this paper, we are only interested in the leading order result that we obtain using the tree level algebra of OPE's between the left invariant currents, which was derived in \BianchiIM\PulettiVB.\foot{The one-loop correction to the classical OPE's have been computed in \BedoyaAV. It would be interesting to use them to compute the normal ordering terms in the antighost \bghost\ and the quantum corrections to the operator ${\cal O}$.} One finds
\eqn\omeg{
{\cal O}=A_0+A_w+A_{ww} \ ,
}
where
\eqn\omegaa{\eqalign{
A_0=&{1\over2\l\lh}\left(\bar J_2^a(J_3\gamma_a KJ_3)-J_2^a (\bar J_3(1-K)\gamma_a J_3)\right)\cr&+{2\over(2\l\lh)^2 }\left(-\bar J_2^a(\l\bar J_3)(\lh\gamma_aJ_3)+J_2^a(\l\bar J_3)(\lh\gamma_a J_3)\right) \ ,
}}
\eqn\omegaw{\eqalign{
A_w=&{1\over (2\l\lh)^2}\Bigl(\half(\l\gamma_a\gamma_{ef}\gamma_b\lh)\bar J_2^aJ_2^b N^{ef}-2(\l\gamma_a\bar J_1)(\lh\gamma_b J_3)N^{ab}\Bigr)\cr&+{1\over 2\l\lh}\Bigl(-(w\gamma_a\gamma_b\l)J_2^a\bar J_2^b-(\l\gamma_a\bar J_1)[w(1-K)\gamma^a J_3]\cr&+[w\gamma^a(1-K)\bar J_3](\l\gamma_aJ_1)-2(\l\bar J_3)(wKJ_1)\Bigr) \ ,
}}
\eqn\omegaww{\eqalign{
A_{ww}=&\half[w(1-K)\gamma_{ef}(1-K)\wh]N^{ef} \ .
}}
The proof that \omeg\ satisfies \hatqb\ is postponed to the Appendix.

\newsec{Type II supergravity background}

In this Section we will 
introduce the action for a generic type II pure spinor superstring sigma model
and show that the $b$ antighost is conserved in the classical BRST cohomology,
in a similar way to the $AdS_5\times S^5$,
whenever the background satisfies certain conditions. 
At the end we will comment on the case when the supergravity background 
does not satisfy such conditions.

The sigma model action for the type II pure spinor superstring in a generic supergravity background
\eqn\curvedaction{\eqalign{
S=&{1\over 2\pi\alpha'}\int d^2z[\half\Pi^a\bar\Pi^b\eta_{ab}+\half \Pi^A\bar \Pi^B B_{AB}+d_\a\bar\Pi^\a+\hat d_\ah\hat{\overline{\Pi}}^\ah) +w_\a\bar\nabla\l^\a+\hat w_\ah\nabla\hat\l^\ah \cr
&+d_\a\hat d_\ah P^{\a\ah}+\l^\a w_\b \hat d_\gh C_\a{}^{\b\gh}+\hat\l^\ah\hat w_\bh d_\g\tilde C_\ah{}^{\bh\g}+
\l^\a w_\b\hat\l^\ah\hat w_\bh S^{\a\ah}{}^{\b\bh}+\alpha'R\Phi(Z) ]
}}
has been studied in \BerkovitsUE, to which we refer the reader for the details. Here we will only describe some features relevant for the present discussion. The worldsheet matter fields are the pullback of the target space super-vielbein $\Pi^A=E^A_M dZ^M$, where $A=(a,\a,\ah)$ is a tangent space superspace index and $M=(m,\mu,\hat\mu)$ a curved superspace index. The ghost content is the same as in the previous case and the covariant derivative on $\l$ ($\hat\l$) is defined using the pullback of the left-moving (right-moving) spin connection $\Omega_\a{}^\b=d Z^M\Omega_{M\a}{}^\b$ ($\hat\Omega_\ah{}^\bh=d Z^M\hat\Omega_{M\ah}{}^\bh$) as
$$
(\nabla\l)^\a=\partial\l^\a+\Omega_\b{}^\a\l^\b \ ,\quad (\nabla\hat\l)^\ah=\partial\hat\l^\ah+\hat \Omega_\bh{}^\ah\hat\l^\bh \ .
$$
The background superfield $B_{AB}$ appearing in \curvedaction\ is the superspace two-form potential; the lowest components of $C_\a{}^{\b\bh}$ and $\tilde C_\ah{}^{\bh\a}$ are related to the gravitini and dilatini; the lowest component of $P^{\a\ah}$ is the Ramond-Ramond bispinor field strength \rrsuperfield; $S_{\a\ah}{}^{\b\bh}$ is related to the Riemann curvature.
The left- and right- moving BRST charges are
\eqn\brst{
Q=\oint d\sigma \l^\a d_\a \ ,\qquad \hat Q=\oint d\sigma\hat\l^\ah \hat d_\ah \ ,
}
where $d$ and $\hat d$ are the pullback of the spacetime supersymmetric derivatives. Conservation of $Q$ and $\hat Q$ and nilpotency of $Q+\hat Q$ imply a set of type IIA/B supergravity constraints, that put the background onshell \BerkovitsUE. It was shown in \BedoyaIC\ that one-loop conformal invariance of the worldsheet action is implied by such constraints. In the following we will recall some of those constraints when needed.

The stress tensor for the pure spinor action in a generic type II supergravity background reads 
\eqn\stressca{
T=-\half\Pi^a\Pi^b\eta_{ab}-d_\a\Pi^\a -w_\a(\nabla\l)^\a \ ,
}

When the R-R superfield $P^{\a\ah}$ is an invertible matrix, we can simplify the sigma model action. The variables $d$ and $\hat d$ couple to the R-R field strength through the term $d_\a\hat d_\ah P^{\a\ah}\to d_\a\hat d^\a$ in the action \curvedaction. If $P$ is invertible we can integrate $d$ and $\hat d$ out upon their equations of motion 
\eqn\dout{\eqalign{
d_\a=&\hat\Pi_\a+\l^\rho w_\sigma P_{\a\ah}C_\rho^{\sigma\ah}\ ,\cr
\hat d^\a=&-\bar\Pi^\a-\hat\l_\rho\hat w^\sigma P_{\a\ah}P^{\rho\hat\rho}P_{\sigma\hat\sigma} \tilde C_{\hat\rho}{}^{\hat\sigma\a} \ .
}}
Substituting \dout\ into the stress tensor \stressca\ we find
\eqn\stresscu{
T= -\half\Pi^a\Pi^b\eta_{ab}-(\hat\Pi_\g+\l^\rho w_\b P_{\g\gh}C_\rho^{\b\gh})\Pi^\g -w_\a(\nabla\l)^\a \ .
}
The proof that the stress tensor \stressca\ is separately  invariant under the BRST transformations generated by the left and right-moving BRST charges
\eqn\stressinv{
\{Q,T\}=\{\hat Q, T\}=0 \ ,
} 
involves the supergravity constraints of \BerkovitsUE\ and is postponed to the Appendix.

\subsec{Some particular backgrounds}

We would like to specialize to a background where the R-R superfield is covariantly constant, namely
\eqn\covac{
\nabla_\a P^{\b\bh}= \nabla_\ah P^{\b\bh}=0 \ .
}
This first requirement on the background ensures that the $b$ antighost we will momentarily introduce satisfies $\{Q,b\}=T$. 

We will also
have a second requirement on the background, namely that the R-R superfield $ P^{\a\ah}$ is such that
\eqn\condP{
 P^{\a\ah}\gamma^k_{\a\b} P^{\b\bh}=f_m^k (\gamma^m)^{\ah\bh}
}
for some tensor superfield $f^k_m$.
By expanding the superfield $P$ in the basis \rrsuperfield\ we can find 
the general solution of the condition \condP\ in terms of the 
R-R $p$-form fluxes. We find that it holds when we turn on a 
single species of $p$-form flux with all nonzero
components sharing $p-1$ legs.
For example, the $p$-form flux 
$$F_{a_1 ... a_P} = \d^1_{[a_1} \d^2_{a_2} ... \d^{p-1}_{a_{p-1}} c_{a_p]}$$
is a solution of \condP\ for any choice of $c_{a_P}$.
In the next subsections, we will show that 
condition \condP\ ensures that $\{\hat Q,b\}=0$ 
and the $b$ ghost is conserved in the BRST cohomology.

When the R-R superfield obeys \condP, we have that $\lh P\gamma^m P\lh=0$ and we can redefine the hatted torsion as $T_{\ah\bh}^a= P_{\alpha\ah}(\g^a)^{\a\b} P_{\b\bh}$. Since the R-R superfield $P^{\a\ah}$ is invertible, by a combined local Lorentz and scale transformation we can reabsorb it into a redefinition of the hatted spinor indices, just as we did in the AdS case in \redefinition, namely
\eqn\redefinitioncurved{\eqalign{
\hat\lambda_\a\equiv P_{\a\hat\a}\hat\l^{\hat\a} \ , &\qquad \hat w^\a\equiv  P^{\a\ah}\hat w_\ah  \ ,\cr
\hat\Pi_\a\equiv  P_{\a\hat\a}\hat \Pi^{\hat\a}\ , &\qquad \hat d^\a\equiv  P^{\a\ah}\hat d_\ah   .}}

When acting on scalar operators  such as the stress tensor and the antighost, the BRST transformations can be cast into the following convenient form
\foot{In
reducing the BRST transformations in the Appendix to the formulas below, one
needs to verify that the contributions of the unhatted and hatted
spin connections in the transformations of the Appendix
cancel independently. This independent cancellation is easily verified 
for the stress tensor and antighost. A similar approach has been previously discussed in \GuttenbergIC\ToninMM. }
\eqn\curvedbrstsca{\eqalign{
 Q \Pi^a=\l\gamma^a\Pi \ , &\qquad \hat Q \Pi^a=\hat\l \gamma^a\hat\Pi  \ ,\cr
 Q \Pi^\a=\nabla\l^\a \ ,&\qquad  \hat Q \Pi^\a=-\hat\l_\b \Pi^a (\gamma_a)^{\b\a} \ ,\cr
 Q \hat\Pi_\b=-\l^\a\Pi^a(\gamma_a)_{\a\b} \ , &\qquad \hat Q \hat\Pi_\a= \nabla\hat\l_\a \ ,\cr
 Q \lambda^\a=Q\hat\l_\a=Q\hat w^\a=0
 \ , &\qquad  \hat Q\l^\a=\hat Q\hat\l_\a=\hat Q w_\b=0
\ ,\cr
Q w_\b =d_\a\ , &\qquad  \hat Q \hat w^\a=\hat d^\a\ ,
 }}
where the background fields $R$ and $\hat R$ are the  Riemann curvatures of the left and right-moving Lorentz connections respectively. 
The BRST transformation of a background tensor superfield is
\eqn\brstfieldssca{\eqalign{
\{Q, \Phi(Z)^{A}_{B}\}=&\l^\a [\nabla_\a \Phi(Z)]^{A}_{B} 
=\lambda^\a\left(\partial_\a \Phi(Z)^{A}_{B}+\Omega_{\a C}{}^{A}\Phi(Z)^{C}_{B}
-\Omega_{\a B}{}^{C}\Phi(Z)^{A}_{C}\right) \ ,\cr
\{\hat Q, \Phi(Z)^{A}_{B}\}=&\hat\l_\a [\nabla^\a \Phi(Z)]^{A}_{B} 
=\hat\lambda_\a\left(P^{\a\ah}\partial_\ah \Phi(Z)^{A}_{B}+P^{\a\ah}\Omega_{\ah C}{}^{A}\Phi(Z)^{C}_{B}
-P^{\a\ah}\Omega_{\ah B}{}^{C}\Phi(Z)^{A}_{C}\right) 
}}
where $\Omega_{\a B}{}^C$ and $\Omega_{\ah B}{}^C$
are hatted or unhatted spin connections depending
if $(B,C)$ are hatted or unhatted spinor indices.

\subsec{Antighost}

As anticipated in the introduction, if the R-R superfield is 
covariantly constant
and invertible, we can follow the same steps as in $AdS_5\times S^5$.
The operator 
\eqn\operatorv{
(\l\lh)\equiv\l^\a P_{\a\ah}\lh^\ah \ ,
}
is in the BRST cohomology and we can
use the inverse of this operator to construct the antighost
\eqn\bghostcurved{
b={1\over 2(\l\hat\l)}(\hat\l\gamma_a\hat\Pi)\Pi^a-w_\a(1-K)^\a_\b \Pi^\b \ , 
}
where we use the curved space projector $K$
\eqn\curvedK{
K_\b^\a={1\over 2(\l\lh)}(\g^a \lh)^\b(\g_a\l)_\a \ ,
}
which satisfies the same properties as in \kappapro\ and is annihilated by both BRST charges.

Let us show that the variation of \bghostcurved\ with respect to the holomorphic BRST charge $Q$  satisfies
$$
\{Q,b\}=T \ ,
$$
where $T$ is given in \stresscu.  The variation of the first term in \bghostcurved\ is
\eqn\qbcurved{
-\half\eta_{ab}\Pi^a\Pi^b -{1\over 2(\l\hat\l)}(\hat\l\gamma_a\hat\Pi)(\l\gamma^a\Pi) \ ,
}
while the second term gives
\eqn\qbcurvedfirst{\eqalign{
Q(-w(1-K)\Pi)=&-\hat\Pi_\b(1-K)^\b_\a\Pi^\a -w_\a(1-K)\nabla\l^\a\ .}}
The BRST holomorphicity constraint
\eqn\holoone{
\nabla_\a P^{\b\gh}+ C_\a^{\b\gh}
=0\ ,
}
together with the condition \covac, implies that $C=0$. 
And the term $-\hat\Pi K \Pi$ in \qbcurvedfirst\ cancels with the second term in \qbcurved\ so we proved that $\{Q,b\}=T$.
%

If the Ramond-Ramond superfield is invertible but is not covariantly constant,
the $b$ ghost of \bghostcurved\ will instead satisfy $\{Q,b\} = T + f_{\a\ah}
(\l^\b \nabla_\b P^{\a\ah})$
for some $f_{\a\ah}$. It should be possible to modify $b \to b - b'$ such that
$\{Q,b'\} = f_{\a\ah}
(\l^\b \nabla_\b P^{\a\ah})$, but we will not attempt to construct $b'$ in this
paper.\foot{We thank Sebastian Guttenberg for pointing out the necessity
of including additional terms in the $b$ ghost when $P^{\a\ah}$ is
not covariantly constant.} 

\subsec{Conservation of the antighost}

We need to show that
\eqn\bzero{
\hat Q b= 0 \ ,
}
so that the argument \showt\ for the conservation of the antighost in the BRST cohomology carries over. We have
\eqn\qhatbcurved{\eqalign{
\hat Q\left({1\over 2(\l\hat\l)}(\hat\l\gamma_a\hat\Pi)\Pi^a\right)=&
{1\over 2(\l\hat\l)}(\hat\l\gamma_a)^\a\left(\nabla\hat\l_\a\Pi^a-\hat\Pi_\a(\hat\l\gamma^a\hat\Pi)\right) \ ,
}}
\eqn\qhatcurvedb{\eqalign{
\hat Q(-w_\b(1-K)^\b_\a\Pi^\a)=&w_\a(1-K)^\a_\b\hat\l^\ah(\gamma_a)_{\ah\bh}P^{\b\bh}\Pi^a\ ,}}
The right hand side of \qhatbcurved\ vanishes on the pure spinor constraint, while the right hand side of \qhatcurvedb\ vanishes due to the properties \kappapro\ of the projector $K$, when the background satisfies \condP. Hence, we proved \bzero\ and the conservation of the antighost up to BRST exact terms.

\subsec{Antighost in a generic type II background}

In this subsection, we discuss the complications
in constructing the antighost in a generic supergravity 
background. If we relax the assumption that the R-R 
superfield $P^{\a\ah}$ be invertible, we cannot integrate out $d_\alpha$ and
$\hat d_{\ah}$  
using their equations of motion \dout. On top of this, we are 
forced to introduce the non-minimal variables as in a flat background. They consist of a new set of left and
right-moving bosonic pure spinors 
$(\bar\l_\a,\hat{\bar\l}_\ah)$ and their conjugate momenta
$(\bar w^\a,\hat{\bar w}^\ah)$, as well as a set of left and
right-moving constrained fermions $(r_\a,\hat r_\ah)$ and their
conjugate momenta $(s^\a,\hat s^\ah)$. 
These non-minimal variables satisfy the constraints
$\bar\l\g^m\bar\l=\hat{\bar\l}
\g^m\hat{\bar\l}=0$ and $\bar\l\g^m r 
= \hat{\bar\l}\g^m\hat r=0$,
so that there are an equal number of non-minimal
bosonic and fermionic degrees of freedom.
After modifying the
BRST operator to include the standard non-minimal term,
these new variables decouple from the cohomology
and the physical spectrum.

The first step in constructing the $b$ antighost in a generic
background would be to find an expression
satisfying $\{Q, b_0\} 
= T$ where
\eqn\firsts{Q = \int dz (\l^\a d_\a +
\bar w^\a r_\a),}
$$T = 
-\half\Pi^a\Pi^b\eta_{ab}-d_\a\Pi^\a -w_\a(\nabla\l)^\a +
s^\a(\nabla r)_\a - \bar w^\a (\nabla \bar\l)_\a \ .$$
Note that in a curved background,
one introduces couplings of the non-minimal variables to
the spin connection in order to make the action
invariant under local Lorentz transformations. This can be done
in a BRST-invariant manner
by adding the BRST-trivial
term 
\eqn\nonminbb{\{Q,- s^\a(\bar\nabla \bar\l)_\a\} = s^\a(\bar\nabla r)_\a
- \bar w^\a(\bar\nabla\bar\l)_\a + \l^\a\bar\Pi^A s^\b \bar\l_\g
R_{\a A \b}{}^\g}
to the minimal action.
Note that
the non-minimal Lorentz
current $\bar N_{ab} = {1\over 2}(-\bar w\g_{ab}\bar\l + s\g_{ab} r)$ is equal to
$\{Q,-{1\over 2}(s\g_{ab}\bar\l)\}$, so the non-minimal action
includes the usual coupling of the spin connection to the Lorentz current. 

The nilpotent BRST transformations on the non-minimal variables
which follow from the curved action are
\eqn\nonminbrst{Q \bar\l_\a = \bar\l_\b (\l^\g\Omega_{\g\a}{}^\b)+
r_\a, }
$$Q \bar w^\a =
\bar w^\b (-\l^\g\Omega_{\g\b}{}^\a)+
\l^\b\l^\g s^\d R_{\b\g\d}{}^\a,$$
$$Q r_\a = r_\b (\l^\g\Omega_{\g\a}{}^\b) + 
\l^\b\l^\g\bar\l_\d R_{\b\g\a}{}^\d,$$
$$Q s^\a = s^\b (-\l^\g\Omega_{\g\b}{}^\a) +\bar w^\a,$$
where the second term in $Q\bar w^\a$ and $Qr_\a$ comes from the 
last term in \nonminbb. 
When acting on scalars, the spin connection $\Omega_{\g\a}{}^\b$ can be dropped
and the non-minimal BRST transformations simplify to
\eqn\nonminbrsttwo{Q \bar\l_\a =
r_\a, \quad Q \bar w^\a = 
\l^\b\l^\g s^\d R_{\b\g\d}{}^\a,}
$$Q r_\a =  
\l^\b\l^\g\bar\l_\d R_{\b\g\a}{}^\d, \quad Qs^\a = \bar w^\a.$$

Using the above non-minimal BRST transformations together with the minimal BRST
transformations of 
\curvedbrstsca, one expects that $b_0$ satisfying
$\{Q,b_0\}=T$ will be a generalization of the
flat-space expression which is
\eqn\completeb{b_0^{flat} =  s^\a\nabla\bar\l_\a  + 
{1\over 2\l\bar\l}(\bar\l\gamma_a)^\a d_\a \Pi^a-
w_\a(\delta^\a_\b-\tilde K_\b^\a)\Pi^\b}
$$+ {{(\bar\l\g^{abc} r)(d\g_{abc} d +24 N_{ab}\Pi_c)}\over{192(\l\bar\l)^2}}
- {{(r\g_{abc} r)(\bar\l\g^a d)N^{bc}}\over{16(\l\bar\l)^3}} +
{{(r\g_{abc} r)(\bar\l\g^{cde} r) N^{ab} N_{de}}\over{128(\l\bar\l)^4}}. $$
However, because the BRST transformations of $d_\a$ and the non-minimal
variables involve the curvature $R_{\a\b\g}{}^\d$, one expects that $b_0$
in a curved background will also have terms depending on this curvature.
Moreover, note that 
the non-minimal version of the projector \projector\ used in \completeb\ is
$$
\tilde K_\a^\b={1\over 2(\l\bar\l)}(\gamma_a\bar\l)^\b(\l\gamma_a)_\a \ .
$$
which has the important difference with the expression
in \bghostcurved\ that the hatted pure spinor has been replaced by
the barred non-minimal pure spinor. 

Because the hatted pure spinor has been replaced with the
non-minimal pure spinor, the right-moving BRST operator
$\hat Q$ is no longer expected to anticommute with $b_0$. Although it will
not be proven here, we conjecture that $\{\hat Q, b_0\} = -\{Q, b_1\}$
for some $b_1$. In other words, we conjecture that its anticommutator
with $\hat Q$ is BRST-trivial with respect to $Q$. Furthermore, we conjecture
that $\{\hat Q, b_1\} = -\{Q, b_2\}$ for some $b_2$, etc.
Note that $b_n$ has left-moving ghost-number $(-1-n)$ and right-moving
ghost-number $n$.

If one assumes this conjecture and
defines $b= b_0 + b_1 + ...$, one finds that 
\eqn\findsco{\{Q+\hat Q, b\}=T.}
Repeating these arguments, one can construct $\hat b= \hat b_0 + \hat b_1 + ...$
such that $\{Q + \hat Q, \hat b\} = \hat T$.
Using the amplitude prescription of \ampli, one can now insert these
composite
$b$ and $\hat b$ antighosts. Although the $b$ and $\hat b$ antighosts
do not have fixed $({\rm left},{\rm right})$ ghost-numbers, the prescription is 
invariant (up to possible surface terms)
under BRST transformations generated by $(Q+\hat Q)$.

As in Ramond-Ramond curved backgrounds, the $b$ antighost
does not necessarily satisfy $\bar\partial b=0$.
In a Ramond-Ramond background, $\hat Q b=0$ implied that 
$\bar\partial b=\{\hat Q,\Omega\}$, which was sufficient
for the consistency of \ampli. However, in a generic curved
background, one needs to use non-minimal variables and $\{\hat Q,b\}$
may be non-zero. Nevertheless, since $\{Q+\hat Q,b\}=T$ and $\bar\partial T=0$,
it might be possible to show that 
$\bar\partial b= \{(Q+\hat Q),\Omega\}$ for some $\Omega$ (and similarly,
$\partial \hat b= \{(Q+\hat Q),\hat\Omega\}$ for some $\hat\Omega$).
If this can be shown, the amplitude prescription of \ampli\ would be consistent, not only for invertible Ramond-Ramond superfields, but for any curved background,
since $Q+\hat Q$ can be pulled off of $\Omega$ and would only generate 
possible surface
terms.

\vskip 15pt
{\bf Acknowledgements:} 
We would especially like to thank S. Guttenberg for pointing out
an error in the first version of the paper that construction
of the $b$ ghost requires additional conditions on $P^{\a\bh}$.
We would also like to thank O. Chandia and
N. Nekrasov for discussions. 
NB would like to thank 
the Simons
Center for Geometry and Physics where part of this research was done
and FAPESP grant 09/50639-2 and
CNPq grant 300256/94-9 for partial financial support.

\appendix{A}{Some results in $AdS_5\times S^5$}

\subsec{Proof of conservation of $b$ in AdS}

Let us check that the BRST transformation of the operator ${\cal O}$ in \hatqb\ is equal to $\bar\nabla b$ in \debi\ and \eomsg.
The BRST transformations of the various terms are 
\eqn\qomega{\eqalign{
\{\hat Q,A_0\}=&C_{33\bar3}+C_{\bar2 3}+C_{23}+C_{2\bar3} \ ,\cr
C_{33\bar3}=&b_0 \ ,\cr
C_{\bar23}=&{2\over2\l\lh}(\lh\gamma_eJ_3)\bar J_{2f}N^{ef}+{2\over(2\l\lh)^2}(\l\gamma_e\gamma_a\lh)(\lh\gamma_fJ_3)\bar J_2^aN^{ef} \ ,\cr
C_{23}=&-\bar\nabla\left({\lh_\a\over 2\l\lh}\right)(\gamma_aJ_3)_\a J_2^a \ ,\cr
C_{2\bar3}=&\half{1\over2\l\lh}\left((\bar J_3\gamma_{ef}\gamma_a\lh)J_2^aN^{ef}-(\bar J_3 K\gamma_{ef}\gamma_a\lh)J_2^aN^{ef}\right)\ ,
}}
\eqn\qomegaw{\eqalign{
\{\hat Q,A_w\}=&B_{2\bar3}+B_{\bar23}+B_1+B_{\bar 1} \ ,\cr
B_{2\bar3}=&\half{1\over2\l\lh}(\bar J_3K\gamma_{ef}\gamma_a\lh)J_2^aN^{ef}-[w(1-K)\gamma_a\bar J_3]J_2^a \ ,\cr
B_{\bar 23}=&[w(1-K)\gamma_b J_3]\bar J_2^b-{2\over2\l\lh}(\l\gamma_a\gamma_c\lh)(\lh\gamma_b J_3)\bar J_2^c N^{ab} \ ,\cr
B_1=&\bar\nabla\left({\lh_\a\over2\l\lh}\right)(w\gamma^a)^\a(\l\gamma_a J_1) \ ,\cr
B_{\bar1}=&-\half(w\gamma_{ef}K\bar J_1)N^{ef}+{2\over2\l\lh}(2\gamma_e\lh)(\l\gamma_f\bar J_1)N^{ef} \ ,
}}
\eqn\qomegaww{\eqalign{
\{\hat Q,A_{ww}\}=&b_{ww}+\half[w\gamma_{ef}K\bar J_1]N^{ef}-{2\over2\l\lh}(w\gamma_e\lh)(\l\gamma_f\bar J_1) N^{ef} \ .
}}
Summing up we find
\eqn\qA{\eqalign{
\{\hat Q,\Omega\}=&b_0+b_w+b_{ww}+\bar\nabla\left({\lh_\a\over2\l\lh}\right)\left(-(\gamma_aJ_3)^\a J_2^a+(w\gamma_a)^\a(\l\gamma^aJ_1)\right) \ ,\cr
=&\bar\nabla b \ .
}}

\appendix{B}{Some results in type IIA/B curved backgrounds}

We consider the case where the R-R superfield is invertible and we have integrated out $d$ upon its equation of motion \dout.
The BRST transformations of the worldsheet fields are generated by the BRST charge $Q+\hat Q$ and we will consider the separate left and right-moving BRST transformations.\foot{The BRST transformations of the heterotic pure spinor superstring in a SYM and SUGRA background have been presented in \ChandiaIX.} We assume the background type II supergravity is onshell and
we use the holomorphicity and nilpotency constraints of \BerkovitsUE\ to simplify the transformations. 
We also use the gauge choice of \BerkovitsUE\ where $T_{\a\b}{}^\g=
T_{\ah\bh}{}^\gh=0 = T_{a \a}{}^\b = T_{a\ah}{}^\bh=0$ and where
$T_{\ah a}{}^b = \hat T_{\a a}{}^b=0$. As explained in
\BerkovitsUE, it is convenient to introduce both
left and right-moving spin connections, $\Omega_{A\b}{}^\g$ and
$\hat\Omega_{A \bh}{}^\gh$ which act respectively on unhatted and
hatted spinor indices. On vector indices, one can use either of these
connections and $T_{A a}{}^b$ is defined using $\Omega_A$ whereas 
$\hat T_{A a}{}^b$ is defined using $\hat\Omega_A$.

The nilpotent BRST transformations are given by:
\eqn\brstfields{
Q \Phi(Z)^{A_1\ldots A_M}_{B_1\ldots B_N}=\l^\a \partial_\a \Phi(Z)^{A_1\ldots A_M}_{B_1\ldots B_N} \ ,\qquad \hat Q\Phi(Z)^{A_1\ldots A_M}_{B_1\ldots B_N}=\hat\l^\ah \partial_\ah \Phi(Z)^{A_1\ldots A_M}_{B_1\ldots B_N}  \ ,
}
\eqn\curvedbrst{\eqalign{
Q Z^M=&\l^\a E_\a^M \ ,\qquad  Q \hat\Pi^\ah=\Pi^\bh(-\l^\g\hat\Omega_{\g\bh}{}^\ah)-\l^\a\Pi^a(\gamma_a)_{\a\b}P^{\b\ah} \ ,\cr
 Q \Pi^a=&\Pi^b(-\l^\g \hat\Omega_{\g b}{}^a)+
\l\gamma^a\Pi \ , \qquad Q \Pi^\a=\Pi^\b(-\l^\g\Omega_{\g\b}{}^\a) +\nabla\l^\a\ ,\cr
 Q d_\g=&-d_\a (-\l^\rho\Omega_{\rho\g}{}^\a)-(\l\gamma_a)_\g\Pi^a + 
\l^\a\l^\b w_\d R_{\g\a\b}{}^\d
 \ ,\cr
 Q \hat d_\gh=&-\hat d_\ah(-\l^\g\hat\Omega_{\g\gh}{}^\ah)+\l^{\hat\rho}\l^\tau\hat w_{\hat\sigma}\hat R_{\gh\tau\hat\rho}{}^{\hat\sigma} \ ,\cr
 Q \lambda^\a=&\lambda^\b(-\l^\g\Omega_{\g\b}{}^\a)
 \ ,\qquad Q w_\b =-w_\a (-\l^\g\Omega_{\g\b}{}^\a)+d_\a\ ,\cr
Q\hat\l^\ah=&\hat\l^\bh(-\l^\g\hat\Omega_{\g\bh}{}^\ah) \ ,\qquad Q\hat w_{\bh}=-w_\ah(-\l^\g\hat\Omega_{\g\bh}{}^\ah) \ .
 }}
\eqn\curvedbrsth{\eqalign{
\hat Q Z^M=&\l^\ah E_\ah^M \ ,\qquad \hat Q \Pi^\a=\Pi^\b(-\hat\l^\gh\Omega_{\gh\b}{}^\a) -\hat\l^\ah \Pi^a (\gamma_a)_{\ah\bh}P^{\a\bh}
 \ , \cr 
\hat Q \Pi^a=&\Pi^b (-\lh^\gh \Omega_{\gh b}{}^a)+
\hat\l\gamma^a\hat\Pi \ ,\qquad  \hat Q \hat\Pi^\ah=\hat\Pi^\bh(-\hat\l^\gh\hat\Omega_{\gh\bh}{}^\ah)+ \nabla\hat\l^\ah \ ,\cr
\hat Q d_\g=&-d_\a(-\hat\l^\gh\Omega_{\gh\g}{}^\a)+\l^\rho\hat\l^{\hat\rho} w_\sigma R_{\g\hat\rho\rho}{}^\sigma \ ,\cr
\hat Q\hat d_\gh=&-\hat d_\ah(-\hat\l^\bh\hat\Omega_{\bh\gh}{}^\ah)+(\hat\l\gamma_a)_\gh\Pi^a +
\lh^\ah\lh^\bh \hat w_\dh \hat R_{\gh\ah\bh}{}^\dh
\ ,\cr
 \hat Q\l^\a=&\l^\b(-\hat\l^\gh\Omega_{\gh\b}{}^\a) \ ,\qquad  \hat Q w_\b=-w_\a(-\hat\l^\gh\Omega_{\gh\b}{}^\a) \ ,\cr
 \hat Q\hat\l^\ah=&\hat\l^\bh(-\hat\l^\gh\hat\Omega_{\gh\bh}{}^\ah) \ ,\qquad  \hat Q \hat w_\bh=-\hat w_\ah(-\hat\l^\gh\hat\Omega_{\gh\bh}{}^\ah)+\hat d_\ah\ .
 }}
The background fields $R$ and $\hat R$ are the  Riemann curvatures of the left and right-moving Lorentz connections $\Omega$ and $\hat\Omega$ respectively. 
In writing the BRST transformation of $\Pi^a$, one can either use the unhatted
or
hatted spin connection. Since $\hat T_{\a a}{}^b=T_{\ah a}{}^b=0$, it
is convenient to use the hatted spin connection in the definition of $Q\Pi^a$
and the unhatted spin connection in the definition of $\hat Q\Pi^a$.
Of course, one can also write $Q\Pi^a$ in terms of the unhatted
spin connection using the relation $\hat\Omega_{\a a}{}^b =
\Omega_{\a a}{}^b - T_{\a a}{}^b$.

We can check nilpotency of these BRST transformations using the supergravity constraints \ChandiaIX. For example, to check that $Q^2=0$, 
\eqn\nilpo{\eqalign{
\epsilon_1 Q (\epsilon_2 Q Z^M)=&\epsilon_1\epsilon_2 \lambda^\b\l^\g\left(\partial_{(\b} E_{\g)}^M+\Omega_{\b\g}{}^\a E_\a^M\right) \cr
=&\epsilon_1\epsilon_2 \lambda^\b\l^\g T_{\b\g}^AE_A^M \ ,
}}
which vanishes because of the torsion constraint $\l^\b\l^\g T_{\b\g}^A=0$. 
We also have
\eqn\nilpol{\eqalign{
\epsilon_1 Q (\epsilon_2 Q \l^\a)=&\epsilon_1\epsilon_2 \lambda^\b\l^\g\l^\rho\left(\partial_{(\rho} \Omega_{\g\b)}{}^\a-\Omega_{(\g\rho}{}^\tau\Omega_{\tau|\b)}{}^\a-\Omega_{(\g\rho}{}^\tau\Omega_{\b)\tau}{}^\a \right) \cr
=&\epsilon_1\epsilon_2 \lambda^\b\l^\g\l^\rho R_{\rho\g\b}{}^\a \ ,
}}
which vanishes due to the constraint $\l^\b\l^\g\l^\rho R_{\b\g\rho}{}^\a=0$.
We can similarly check that the supergravity constraints imply
that $\hat Q^2=0$ and $\{Q,\hat Q\}=0$.

When acting on a target space scalar operator, we can rearrange the BRST transformations by removing the Lorentz spin connection
from the transformation of the worldsheet fields \curvedbrst\ and \curvedbrsth\ and covariantizing the BRST transformations of the background superfields \brstfields\ as
\eqn\brstfieldssca{\eqalign{
\{Q, \Phi(Z)^{A}_{B}\}=&\l^\a [\nabla_\a \Phi(Z)]^{A}_{B} 
=\lambda^\a\left(\partial_\a \Phi(Z)^{A}_{B}+\Omega_{\a C}{}^{A}\Phi(Z)^{C}_{B}
-\Omega_{\a B}{}^{C}\Phi(Z)^{A}_{C}\right) \ ,\cr
\{\hat Q, \Phi(Z)^{A}_{B}\}=&\hat\l^\ah [\nabla_\ah \Phi(Z)]^{A}_{B} 
=\hat\lambda^\ah\left(\partial_\ah \Phi(Z)^{A}_{B}+\Omega_{\ah C}{}^{A}\Phi(Z)^{C}_{B}
-\Omega_{\ah B}{}^{C}\Phi(Z)^{A}_{C}\right) .
}} 
Using these redefined transformations, one can check that the stress tensor is BRST invariant.

When the R-R superfield satisfies \condP, we can further redefine the worldsheet fields as in \redefinitioncurved\ and finally obtain \curvedbrstsca\ and \brstfieldssca.
The only subtlety is that both the unhatted and hatted spin connections
appear in \curvedbrst\ and \curvedbrsth, so one needs to verify that
they cancel independently in the transformation of the scalar operator.
Fortunately, this is easily verified for the stress tensor and
antighost of \stressca\ and
\bghostcurved. The unhatted and hatted spin connections
appearing in the BRST transformations of unhatted and hatted spinor fields
are easily shown to cancel. And the hatted spin connection appearing in
the BRST transformation $Q\Pi^a$ cancels since $\Pi^a$ only appears
in the combinations $\eta_{ab}\Pi^a\Pi^b$ and $(\lh\g_a\hat\Pi)\Pi^a$
in \stressca\ and \bghostcurved.

\subsec{BRST invariance of the stress tensor}

Let us check that the stress tensor \stresscu\ is BRST invariant. Since we are acting on a scalar operator, we can use the redefined BRST transformations \brstfieldssca. First consider the left-moving BRST variation $\{Q,T\}$. We find 
\eqn\qtzeroone{\eqalign{
Q\left(-\half\Pi^a\Pi^b\eta_{ab}\right)=& -\eta_{ab}\Pi^a(\l\gamma^b\Pi) \ ,}}
\eqn\qtzero{\eqalign{
Q\left(-P_{\g\gh}\hat\Pi^\gh\Pi^\g\right)=&-\l^\rho(\nabla_{\rho}P_{\g\gh})\hat\Pi^\gh\Pi^\g+
\l^\a\Pi^a(\gamma_a)_{\a\rho}\Pi^\rho+P_{\g\gh}\hat\Pi^\gh\nabla\l^\g \ ,}}
\eqn\qtzerofir{\eqalign{
Q\left(-P_{\g\gh}\l^\rho w_\b C_\rho^{\b\gh}\Pi^\g\right)=&
-\l^\rho(\nabla_\rho P_{\g\gh})\l^\a w_\b C_\a^{\b\gh} \Pi^\g-
P_{\g\gh}\l^\a\l^\l w_\sigma C_{\l}^{\sigma\bh}C_\a^{\b\gh}\Pi^\g \cr
&-P_{\g\gh}\l^\a\hat\Pi^\bh C_\a^{\b\gh}\Pi^\g \cr
&-P_{\g\gh}\l^\a w_\b \l^\rho\nabla_\rho C_\a^{\b\gh}\Pi^\g\cr
&+P_{\g\gh}\l^\a w_\b C_\a^{\b\gh}\nabla\l^\g \ ,
}}
\eqn\qtzerosec{\eqalign{
Q\left(-w_\a\nabla\l^\a\right)=&-P_{\a\gh}(\hat\Pi^\gh+\l^\sigma w_\rho C_\sigma^{\rho\gh})\nabla\l^\a+w_\b \l^\a\Pi^\rho\l^\delta R_{\rho\delta\a}{}^\b \ ,
}}
where in the last equation we used the fact that $\Omega_\a{}^\b=\Pi^A\Omega_{A\a}{}^\b$ and
\eqn\qomegat{
Q\Omega_\a{}^\b=\Pi^A\l^\gamma R_{A\g\a}{}^\b \ ,
}
and $\l^\g\l^\b R_{a\g\b}{}^\rho=\l^\g\l^\b R_{\ah\g\b}{}^\rho=0$ from the BRST nilpotency contraints. Let us simplify the previous expressions, 
noting that
\eqn\psimpli{
\nabla_\rho P_{\g\gh}=-P_{\g\ah}(\nabla_\rho P^{\a\ah})P_{\a\gh} \ .
}
Due to the holomorphicity constraint of \holoone\
the first line in \qtzerofir\ vanishes, while the first term in \qtzero\
 cancels against the second line in \qtzerofir.
The last term in \qtzero\ cancels against the last term in \qtzerofir\ plus the first term in \qtzerosec. 
Finally, using the BRST holomorphicity constraint $\l^\a\l\b(\nabla_\a C_\b^{\g\gh}-P^{\delta\gh}R_{\rho\a\b}{}^\g)=0$, the third line in \qtzerofir\ cancels against the last term in \qtzerosec. Hence the result  $\{Q,T\}=0$.

Let us check that the right-moving BRST variation vanishes as well. The various terms in \stressca\ transform as
\eqn\hatqT{\eqalign{
\hat Q \left(-\half\Pi^a\Pi^b\eta_{ab}\right)=&-
\eta_{ab}\Pi^a(\hat\l\gamma^b\hat\Pi) \ ,}}
\eqn\hatqTfir{\eqalign{
\hat Q\left(-P_{\g\gh}\Pi^\g\hat\Pi^\gh\right)=&-\hat\l^{\hat\rho}\nabla_{\hat\rho}P_{\g\gh}\Pi^\g\hat\Pi^\gh+P_{\g\gh}\Pi^\g\nabla\hat\l^\ah \cr
&\hat\l^{\hat\sigma}\Pi^a(\gamma_a)_{\hat\sigma\gh}\hat\Pi^\gh \ ,
}}
\eqn\hatqTsec{\eqalign{
\hat Q\left(-P_{\g\gh}\l^\rho w_\b C_\rho^{\b\gh}\Pi^\g\right)=&-\hat\l^{\hat\rho}\nabla_{\hat\rho}P_{\g\gh}\l^\a w_\b C_{\a}^{\b\gh} \Pi^\g\cr
&-P_{\g\gh}\l^\a w_\b\hat\l^{\hat\rho}\nabla_{\hat\rho} C_\a^{\b\gh}\Pi^\g\cr
&-\l^\a w_\b C_\a^{\b\gh}\hat\l^{\hat\rho}\Pi^a(\gamma_a)_{\hat\sigma\gh} \ ,
}}
\eqn\hatqTthir{\eqalign{
\hat Q(-w_\a\nabla\l^\a)=&-w_\a\hat\l^\ah\l^\b\Pi^a R_{a\ah\b}{}^\a-w_\a\hat\l^\ah\l^\b\Pi^\g R_{\g\ah\b}{}^\a \ .
}}
The first term \hatqT\ cancels against the second line of \hatqTfir. Using the BRST holomorphicity constraint 
$$\nabla_\ah C_\b^{\g\hat\delta}-P^{\rho\hat\delta}R_{\rho\ah\b}{}^\g-S_{\b\ah}{}^{\g\hat\delta} =0\ ,
$$
we can recast the second line of \hatqTsec\ into
\eqn\qlinesec{
-P_{\g\gh}\l^\a w_\b\hat\l^{\hat\rho}\nabla_{\hat\rho} C_\a^{\b\gh}\Pi^\g=\l^\a w_\b\hat\l^\ah \Pi^\g R_{\g\ah\a}{}^\b+\l^\a w_\b\hat\l^\ah \Pi^\g P_{\g\gh}S_{\a\hat\rho}{}^{\b\gh} \ ,
}
then we see that the first term in \qlinesec\ cancels against the last term in \hatqTthir\ while using the holomorphicity constraint $R_{a\ah\b}{}^\g=C_\b^{\g\hat\delta}T_{\hat\delta\ah a}$ we find that the first term in \hatqTthir\ cancels against the last line in \hatqTsec.
The remaining terms vanish due to the equations of motion for $\hat\l$ in a curved background
\eqn\eomlh{
\eqalign{
\nabla\hat\l^\ah=&-\hat\l^\bh P_{\g\gh}(\hat\Pi^\gh+\l^\rho w_\sigma C_\rho^{\sigma\gh}) \tilde C_{\bh}^{\ah \g}-\hat\l^\bh\l^\a w_\b S_{\a\bh}{}^{\b\ah} \ ,
}}
by using the holomorphicity constraint $\tilde C_{\ah}^{\gh\b}-\nabla_\ah P^{\b\gh}=0$ in the gauge $P^{\b\hat\delta} T_{\hat\delta\ah}{}^\gh=0$. Hence we proved that $\{\hat Q, T\}=0$.

\listrefs

\bye